\documentclass[english]{iopart}
\usepackage[T1]{fontenc}
\usepackage[latin9]{inputenc}
\usepackage{units}
\usepackage{amstext}
\usepackage{amssymb}
\usepackage{graphicx}
\usepackage{esint}

\makeatletter
\usepackage{iopams}
\usepackage{setstack}

\makeatother

\usepackage{babel}
\begin{document}

\title{The Nonlinear Schr\"{o}dinger Equation with a random potential: Results
and Puzzles}

\author{Shmuel Fishman}

\address{Physics Department, Technion - Israel Institute of Technology, Haifa
32000, Israel.}

\author{Yevgeny Krivolapov}

\address{Physics Department, Technion - Israel Institute of Technology, Haifa
32000, Israel.}

\author{Avy Soffer}

\address{Mathematics Department, Rutgers University, New-Brunswick, NJ 08903,
USA.}
\begin{abstract}
The Nonlinear Schr\"{o}dinger Equation (NLSE) with a random potential
is motivated by experiments in optics and in atom optics and is a
paradigm for the competition between the randomness and nonlinearity.
The analysis of the NLSE with a random (Anderson like) potential has
been done at various levels of control: numerical, analytical and
rigorous. Yet, this model equation presents us with a highly inconclusive
and often contradictory picture. We will describe the main recent
results obtained in this field and propose a list of specific problems
to focus on, that we hope will enable to resolve these outstanding
questions.
\end{abstract}
\maketitle

\section{Introduction}

The Nonlinear Schr\"{o}dinger Equation (NLSE) with a random potential
is a fundamental problem. In spite of extensive mathematically rigorous,
analytical and numerical explorations, the elementary properties of
its dynamics are not known. The problem is relevant for experiments
and its resolution will shed light on many problems in chaos and nonlinear
physics. It may also stimulate novel experiments. On a one-dimensional
lattice, the NLSE with a random potential (that will be the subject
of the present review) is given by,
\begin{equation}
i\partial_{t}\psi\left(x,t\right)=H_{0}\psi\left(x,t\right)+\beta\left\vert \psi\left(x,t\right)\right\vert ^{2}\psi\left(x,t\right),\label{eq:NLSE}
\end{equation}
 where
\begin{equation}
H_{0}\psi\left(x,t\right)=-J\left[\psi\left(x+1,t\right)+\psi\left(x-1,t\right)\right]+\varepsilon_{x}\psi\left(x,t\right),\label{eq:Lin_Hamiltonian}
\end{equation}
 while, $x\in\mathbb{Z};$ and $\left\{ \varepsilon_{x}\right\} $
is a collection of i.i.d. random variables uniformly distributed in
the interval $\left[-\frac{w}{2},\frac{w}{2}\right]$. The Hamiltonian
$H_{0}$ is the Anderson model in one-dimension \cite{Anderson1958,Ishii1973,Abrahams1979,Lee1985,Lifshits1988}.
It is important to note that for the dynamics generated by (\ref{eq:NLSE})
the $\ell^{2}$ norm, $\mathcal{N}=\sum_{x}\left|\psi\left(x,t\right)\right|^{2}$,
and the energy given by the Hamiltonian (\ref{eq:NLS_Hamiltonian})
are conserved \cite{Sulem1999}.

In the present review we will focus on the question about the dynamics;
will an initial wave function $\psi\left(x,t=0\right)$, which is
localized in space, spread indefinitely for large times, and in particular
in the asymptotic limit, $t\to\infty$. Surprisingly, the answer to
this elementary question is not known in spite of extensive research
in the last two decades \cite{Bourgain2007,Pikovsky2008,Wang2008,Flach2009,Flach2009a,Skokos2009,Fishman2009a,Krivolapov2010}.
We believe that the NLSE (\ref{eq:NLSE}) is a representative of many
nonlinear problems, such as the famous Fermi-Pasta-Ulam (FPU) problem
\cite{Berman2005}. Therefore, its understanding may shed light on
dynamics generated by other nonlinear equations, e.g, nonlinear Klein-Gordon
and FPU equations. Many properties of (\ref{eq:NLSE}) will be shared
by the continuous counterpart, where $x$ is a continuous variable.
Since most of the results that were derived so far are for the discrete
problem, in the present review we will confine ourselves to this case.
The dynamics is completely understood in the two limiting cases. In
the absence of the random potential ($\varepsilon_{x}=0$, for all
$x$) an initially localized wavepacket will spread indefinitely,
for all values of $\beta$, unless solitons are formed. In the discrete
case, unlike the continuous case, the formation of the solitons cannot
be established rigorously \cite{Bronski1998}. The continuous version
of this model is in fact an integrable problem \cite{Sulem1999}.
For attractive nonlinearity, $\beta<0$, solitons are found while
for repulsive nonlinearity, $\beta>0$, complete spreading takes place.
In the presence of randomness $\left(w>0\right)$, but for $\beta=0$
Eq. (\ref{eq:NLSE}) reduces to the Anderson model (\ref{eq:Lin_Hamiltonian})
where it is rigorously established that \emph{all} the eigenstates
are exponentially localized in one-dimension with probability one
\cite{Anderson1958,Ishii1973,Lee1985,Lifshits1988}. At long scales
the eigenfunctions behave as
\begin{equation}
u_{n}\left(x\right)\sim e^{-\left|x-x_{n}\right|/\xi},
\end{equation}
 where $x_{n}$ is the localization center and $\xi$ is the localization
length. Consequently, diffusion is suppressed and in particular a
wavepacket that is initially localized will not spread to infinity.
This is the phenomenon of Anderson localization. In two-dimensions
it is known heuristically from the scaling theory of localization,
that all the states are localized, while in higher dimensions there
is a mobility edge that separates localized and extended states \cite{Abrahams1979,Lee1985}.

The behavior of the dynamics generated by (\ref{eq:NLSE}) is very
different in the two extreme limits $\left(w=0,\,\beta\neq0\right)$
and $\left(w\neq0,\,\beta=0\right)$. Therefore, it is a paradigm
for the exploration of the competition between randomness and nonlinearity.
Let us comment on the choice of the nonlinearity.

The nonlinear term of the form $\beta\left|\psi\right|^{2}\psi$ used
in (\ref{eq:NLSE}) is just one possibility, which is used in this
review for the sake of clarity. In several theoretical studies it
is replaced by \cite{Pikovsky2008,Veksler2009,Flach2009,Mulansky2009}
\begin{equation}
H_{\sigma}=\beta\left|\psi\right|^{2\sigma}\psi,\label{eq:Gen_Nonlin_term}
\end{equation}
 where $\sigma>0$ is arbitrary. In some mathematical studies it is
also replaced by $H_{\beta\left(x\right)}=\beta\left(x\right)\left|\psi\right|^{2}\psi$,
where $\beta\left(x\right)$ is a decaying function of $x$ \cite{Bourgain2007}.
Other types types of nonlinear terms appear in experimental realizations.

The NLSE was derived for a variety of physical systems under some
approximations. It was derived in classical optics, where $\psi$
is the electric field, by expanding the index of refraction in powers
of the electric field, keeping only the leading nonlinear term \cite{Agrawal2007}.
Let the index of refraction depend on the intensity of the electric
field $\left|\psi\right|^{2}$, then for weak fields it takes the
form,
\begin{equation}
n\left(\left|\psi\right|^{2}\right)=n_{0}+n_{1}\left|\psi\right|^{2}+O\left(\left|\psi\right|^{4}\right).\label{eq:Nonlin_Refrac_indx}
\end{equation}
 The nonlinear term in (\ref{eq:NLSE}) corresponds to a weak field
so that the quartic correction is negligible. In several important
cases $n\left(\left|\psi\right|^{2}\right)$ saturates, namely, $\lim_{\left|\psi\right|^{2}\to\infty}n\left(\left|\psi\right|^{2}\right)=\text{const}$.
For example in the induction technique \cite{Efremidis2002,Fleischer2003}
the index of refraction takes the form,
\begin{equation}
n\left(\left|\psi\right|^{2}\right)=\frac{n_{0}}{1+\left|\psi\right|^{2}}.
\end{equation}

In optics, Eq. (\ref{eq:NLSE}) corresponds to the paraxial approximation
where the propagation direction plays the role of time. In this approximation
the variation in the index of refraction in space is weak, and therefore
there is only a small change in the propagation direction, and back-scattering
is negligible.

For Bose-Einstein Condensates (BEC), the NLSE is a mean field approximation,
where the term proportional to the density $\beta|\psi|^{2}$ approximates
the interaction between the atoms. In this field the NLSE is known
as the Gross-Pitaevskii Equation (GPE) \cite{Pitaevskii1961,Gross1961,Gross1963,Dalfovo1999,Leggett2001,Pitaevskii2003}.
It was rigorously established, for a large variety of interactions
and of physical conditions, that the NLSE (or the GPE) is exact in
the thermodynamic limit \cite{Erdos2007,Lieb2002}. Experiments on
spreading of wavepackets of cold atoms in a random optical potential
were recently performed \cite{Clement2005,Lye2005,Clement2006,Sanchez-Palencia2007}.
In those experiments as in experiments in optics, the random potential
exhibits correlations and therefore deviates from the model presented
in (\ref{eq:NLSE}).

Another possible form of the nonlinear term is
\begin{equation}
H_{H}=\psi\left(x\right)\int V\left(x-x'\right)\left|\psi\left(x'\right)\right|^{2},\label{eq:nonlocal_nonlin}
\end{equation}
 which results in the Hartree approximation extensively used in solid-state
physics. The Gross-Pitaevskii equation (or NLSE) is obtained from
(\ref{eq:nonlocal_nonlin}) in the limit where $\left|\psi\left(x'\right)\right|^{2}$
is slowly varying.

The theory of Anderson localization was very recently extended to
the many-body particle systems \cite{Aizenman2008,Basko2006,Basko2007}.
It was found that indeed for sufficiently low energies localization
takes place for fermions \cite{Basko2006,Basko2007} as well as for
bosons \cite{Aleiner2010}. It should be emphasized that in these
works localization is analyzed for the case where the density is non-vanishing
in the thermodynamic limit. The problem of spreading with a vanishing
average density, corresponding to the problem that is the subject
of the present review, is different in principle in the thermodynamic
limit. For the $N$ body problem, a wavepacket that is initially localized
will remain localized, as established rigorously in \cite{Aizenman2009,Chulaevsky2009}
(see Subsection \ref{sub:Many-body-localization}).

The model (\ref{eq:NLSE}) was motivated by experimental realizations
we have discussed above, but this review will treat the problem of
spreading, and in particular the asymptotic one, as a fundamental
theoretical problem.

For linear problems, all aspects of dynamics are determined by the
spectral properties, namely the eigenvalues and the eigenfunctions.
This is not correct for nonlinear problems. For example, for small
$\beta$ in (\ref{eq:NLSE}) there are stationary and quasi-periodic
states which are exponentially localized \cite{Albanese1988,Albanese1991,Bourgain2008}.
This however does \emph{not} imply that an initially localized wavepacket
will not spread, contrary to the case of a linear system with a bounded
localization length. Transmission through a chain (\ref{eq:NLSE})
was extensively studied \cite{Doucot1987,Doucot1987a,Hennig1999,Paul2007},
but since it is not directly related to the spreading problem (unlike
the situation for linear systems), it will be left out of the scope
of the present review.

The natural question we will survey is whether a wavepacket, that
is initially localized in space, will indefinitely spread for dynamics
controlled by (\ref{eq:NLSE}). A simple heuristic argument indicates
that spreading will be suppressed by randomness. If unlimited spreading
takes place, the amplitude of the wave function will decay since the
$\ell^{2}$ norm, $\mathcal{N}$, is conserved. Consequently, the
nonlinear term will eventually become negligible, and Anderson localization
will take place as a result of the randomness, as was conjectured
by Fr\"{o}hlich \emph{et al }\cite{Frohlich1986}. However, in numerical
calculations performed by Shepelyansky \cite{Shepelyansky1993} for
the kicked rotor with a cubic nonlinear term, Anderson localization
(that takes place in the absence of the nonlinear term) was destroyed
and sub-diffusion takes place. Similar spreading was found numerically
also by Shepelyansky and Pikovsky \cite{Pikovsky2008} and by Flach
and coworkers \cite{Skokos2009,Flach2009}. Therefore, the naive argument
for localization of (\ref{eq:NLSE}) has to be reconsidered and a
proper theory should be developed. A natural question is what can
we conclude from the numerical simulations ? The main problem is that
dynamics of (\ref{eq:NLSE}) are chaotic. The dynamics are generated
by the Hamitonian,
\begin{eqnarray}
H_{\text{NLS}}\left(\psi\left(x,t\right)\right) & = & \sum_{x}\left[J\left(\psi\left(x+1\right)\psi^{*}\left(x\right)+\psi^{*}\left(x+1\right)\psi\left(x\right)\right)\right.\label{eq:NLS_Hamiltonian}\\
 & + & \left.\varepsilon_{x}\left|\psi\left(x\right)\right|^{2}+\frac{\beta}{2}\left|\psi\left(x\right)\right|^{4}\right]\nonumber
\end{eqnarray}
 Where the NLSE (\ref{eq:NLSE}) is the corresponding Hamilton's equation
with the conjugate variables $\psi\left(x\right)$ and $\psi^{*}\left(x\right)$.
Due to the nonlinearity, the motion in the $\psi\left(x\right)$,$\psi^{*}\left(x\right)$
phase-space will be typically chaotic. Therefore, the numerical solutions
of (\ref{eq:NLSE}) are not the actual solutions. In order to draw
conclusions it is assumed that they are statistically similar to the
correct solutions. Since it is a system of an infinite number of degrees
of freedom there is no real theoretical support for this assumption.
If we use the fact that only a finite number of the $\psi\left(x\right)$
variables are involved, there is a competition between two effects.
Chaos is enhanced by increase in effective number of degrees of freedom,
and suppressed by the decreasing amplitude of the spreading wavepacket.
This competition is outlined in Subsection \ref{sub:Scaling}. There
may be also technical problems with the numerical algorithm, this
will be discussed in Section \ref{sec:Numerics}.

Another model similar to (\ref{eq:NLSE}), that was extensively studied,
is the quartic Klein-Gordon equation \cite{Skokos2009,Flach2009},
\begin{equation}
H_{\text{KG}}=\sum_{x}\frac{1}{2}p_{x}^{2}+\frac{1}{2}\varepsilon_{x}q_{x}^{2}+\frac{1}{4}q_{x}^{4}+\frac{1}{w}\left(q_{x+1}-q_{x}\right)^{4}.
\end{equation}

This model differs from the NLSE with a random potential, where the
modes of the linear problem are effectively localized on a few lattice
sites. The perturbation theory is well controlled and Nekhoroshev
type estimates were established \cite{Benettin1988}.

A central problem in the field is the Fermi-Pasta-Ulam (FPU) problem
\cite{Berman2005}. It is interesting to point out that for this problem
the spreading starts after a long time, although it is very different
from the NLSE and the nonlinear Klein-Gordon problems. For the NLSE
and the quartic Klein-Gordon equations, only neighboring modes in
space, of the corresponding linear problem are coupled, while for
the FPU problem, all the modes are coupled.

\section{\label{sec:Theoretical-analysis}Theoretical analysis}

In this section various non rigorous theories for the spreading mechanism
will be presented. The purpose is to analyze the various regimes starting
from the short time regime and up to the asymptotic time regime. Various
authors use an expansion of the wavefunction in terms of the eigenstates,
$u_{m}\left(x\right)$, and eigenvalues, $E_{m}$, of $H_{0}$ as,
\begin{equation}
\psi\left(x,t\right)=\sum_{m}c_{m}\left(t\right)e^{-iE_{m}t}u_{m}\left(x\right).\label{eq:expansion}
\end{equation}
 For the nonlinear equation the dependence of the expansion coefficients,
$c_{n}\left(t\right),$ is found by inserting this expansion into
(\ref{eq:NLSE}), resulting in
\begin{equation}
i\partial_{t}c_{n}=\beta\sum_{m_{1},m_{2},m_{3}}V_{n}^{m_{1}m_{2}m_{3}}c_{m_{1}}^{\ast}c_{m_{2}}c_{m_{3}}e^{i\left(E_{n}+E_{m_{1}}-E_{m_{2}}-E_{m_{3}}\right)t}\label{eq:c_n_exact}
\end{equation}
 where $V_{n}^{m_{1}m_{2}m_{3}}$ is an overlap sum
\begin{equation}
V_{n}^{m_{1}m_{2}m_{3}}=\sum_{x}u_{n}\left(x\right)u_{m_{1}}\left(x\right)u_{m_{2}}\left(x\right)u_{m_{3}}\left(x\right).\label{eq:overlap_int}
\end{equation}
 This sum is negligibly small if the various eigenfunctions are not
localized in the same region of the order of the localization length,
$\xi$.

The eigenvalues $E_{n},$ the eigenfunctions\c{}$u_{n}\left(x\right),$
the expansion coefficients, $c_{n}\left(t\right)$ and the overlap
sums depend on the random potentials, $\left\{ \varepsilon_{x}\right\} $
and therefore they are \emph{random varibales}. Consequently, $E_{n}$,
$u_{n}\left(x\right)$ and $V_{n}^{m_{1}m_{2}m_{3}}$ take different
values for the various realizations of the potentials, $\left\{ \varepsilon_{x}\right\} $.
In Subsections \ref{sub:Effective-noise-theories} and \ref{sub:Scaling}
phenomenological theories are discussed, and in Subsection \ref{sub:Chaotic-spot-model}
a relation to phase-space structures is briefly presented, while in
Subsection \ref{sub:The-renormalized-perturbation} a systematic perturbation
theory is developed.

\subsection{\label{sub:Effective-noise-theories}Effective noise theories}

\begin{figure}
\begin{centering}
\includegraphics[width=8cm]{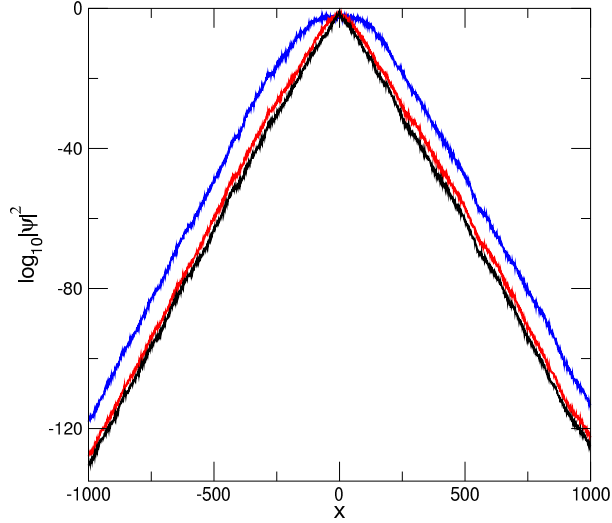}
\par\end{centering}

\caption{\label{fig:Shapo}(color online) Probability distribution $\left|\psi\right|^{2}$
over lattice sites $x$ for $J=1$, $w=4$ and for $\beta=1$, $t=10^{8}$
(top blue/solid curve); $t=10^{5}$ (middle red/grey curve); $\beta=0$,
$t=10^{5}$ (bottom black curve). (Fig. 2 of \cite{Pikovsky2008}).}
\end{figure}

In this subsection we present the phenomenological theory \cite{Skokos2009,Flach2009}
for the spreading that is found numerically and will be described
in detail in Section \ref{sec:Numerics}. It is clear that not all
the content of the initial wavepacket spreads \cite{Kopidakis2008}.
It was rigorously shown, that for sufficiently large $\beta$, the
initial wavepacket cannot spread so that its amplitude everywhere
vanishes at infinite time \cite{Kopidakis2008}. It does not contradict
spreading of a fraction of the wavepacket. For a wavepacket initially
localized on one lattice site or started at one linear eigenstate
of $H_{0}$, sub-diffusion was found in numerical experiments \cite{Shepelyansky1993,Molina1998,Kopidakis2008,Flach2009,Skokos2009}.
The purpose of the theory presented in what follows, is to explain
the spreading that takes place after some time. It was found numerically
that after some initial time the shape of the wavepacket is similar
to the one found in Fig \ref{fig:Shapo}.

\begin{figure}
\centering{}\includegraphics[width=8cm]{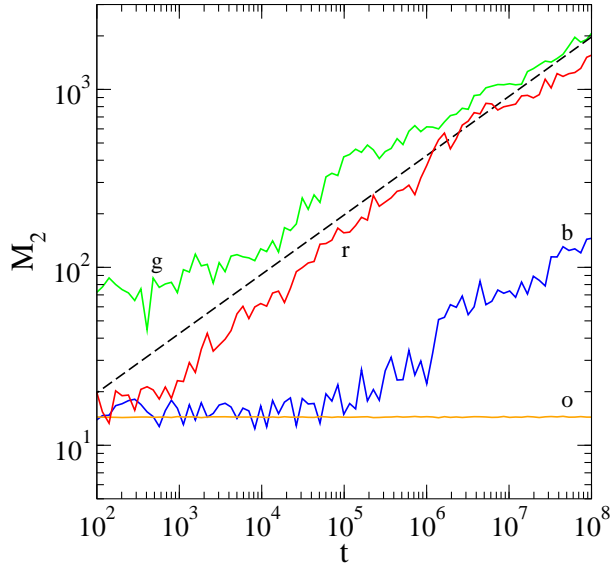} \caption{\label{fig1}(color online) $M_{2}$ versus time in log-log plots.
For $J=1$, $w=4$ and $\beta=0,0.1,1,4.5$ ((o)range, (b)lue, (g)reen,
(r)ed). The disorder realization is kept unchanged. The dashed straight
line guides the eye for exponent 1/3 (Fig. 2 of \cite{Skokos2009}).}
\end{figure}

It consists of a relatively flat region at the center and exponentially
decaying tails. The theory of \cite{Skokos2009,Flach2009} assumes
spreading from the relatively flat region of the wavepacket to the
region where the amplitude of the wavepacket is small. Let $m_{1},m_{2}$
and $m_{3}$ designate eigenstates of $H_{0}$ with the centers of
localization found within the flat region, and let $n$ designate
a state with a center of localization found in the tail of the wavepacket,
but in the vicinity of the flat region. Therefore, spreading will
take place to the region where the $n$-th state is localized,
\begin{equation}
|c_{m_{1}}|^{2}\approx|c_{m_{2}}|^{2}\approx|c_{m_{3}}|^{2}\approx\rho\label{cm}
\end{equation}
 while
\begin{equation}
|c_{n}|^{2}\ll\rho.\label{cn}
\end{equation}
 It is further assumed that the RHS of (\ref{eq:c_n_exact}) is a
random function denoted by $F_{n}\left(t\right)$, of the form \cite{Skokos2009},
\begin{equation}
F_{n}=C_{1}P\beta\rho^{3/2}f_{n}(t)\label{F}
\end{equation}
 where
\begin{equation}
P=C_{2}\beta\rho.\label{eq:P_Flach}
\end{equation}
 Equation (\ref{eq:P_Flach}) is the probability of a {}``resonance''
between four modes, $C_{1}$ and $C_{2}$ are constants, while $f_{n}\left(t\right)$
is a random function with a rapidly decaying correlation function,
$C\left(t\right)$. Introduction of $P$ and the assumption that it
satisfies (\ref{eq:P_Flach}), in particular its linearity in $\beta$,
are the strongest assumptions of the theory, which still require justification.
In \cite{Skokos2009} it is claimed that numerical calculations support
this assumption. Under these assumptions (\ref{eq:c_n_exact}) reduces
to
\begin{equation}
i\partial_{t}c_{n}(t)=F_{n}(t).\label{eq:c_n-F_n}
\end{equation}
 Assuming that $F_{n}(t)$ can be considered random, with rapidly
decaying correlations in time, integration results in
\begin{equation}
c_{n}(t)=-iC_{1}P\beta\rho^{3/2}\int^{t}dt'f_{n}(t').
\end{equation}
 Averaging over realizations one finds
\begin{equation}
\left\langle \left|c_{n}\right|^{2}\right\rangle =C_{3}\beta^{4}\rho^{5}t,\label{eq:Flach_amp_growth}
\end{equation}
 where
\begin{equation}
C_{3}=C_{1}C_{2}\int_{0}^{\infty}C\left(t'\right)\,\mathrm{d}t'.\label{eq:Correlation_function}
\end{equation}
 The equilibration time, $T$, is the time when $\left\langle \left|c_{n}\right|^{2}\right\rangle \approx\rho$
,
\begin{equation}
T\propto\frac{1}{\beta^{4}\rho^{4}}.
\end{equation}
 During the time, $T$, the flat region spreads so that it covers
the site where the state $n$ was localized. The resulting diffusion
coefficient satisfies,
\begin{equation}
D\propto\frac{1}{T}\propto\beta^{4}\rho^{4}.\label{eq:diffusion coefficient}
\end{equation}
 At time scales $t\gg T$, diffusion takes place and
\begin{equation}
\frac{1}{\rho^{2}}\propto M_{2}\propto Dt\propto\beta^{4}\rho^{4}t\label{eq:ro_D_t}
\end{equation}
 Where $M_{1}=\sum x\left|\psi\left(x,t\right)\right|^{2}$ is the
first moment and
\begin{equation}
M_{2}=\sum_{x}\left(x-M_{1}\right)^{2}\left|\psi\left(x,t\right)\right|^{2}
\end{equation}
 is the second moment. Therefore,
\begin{equation}
\frac{1}{\rho^{2}}\propto\left(\beta^{4}t\right)^{1/3},\label{eq:1 over ro_sq}
\end{equation}
 and the second moment satisfies,

\begin{equation}
M_{2}\propto\beta^{4/3}t^{1/3}.\label{eq:m2-1}
\end{equation}
 In agreement with the numerical results presented in Fig. \ref{fig1}.

The equilibration time satisfies,

\begin{equation}
T\propto\beta^{-4/3}t^{2/3}.
\end{equation}
 Therefore the theory is consistent, as $T\ll t$ for large $t$.

The crucial assumption of the theory presented in this section is
that $F_{n}\left(t\right)$ behaves as noise with a rapidly decaying
correlation function so that the integral (\ref{eq:Correlation_function})
converges. This assumption was explicitly tested \cite{Michaely2011}.
The reasons for $F_{n}\left(t\right)$ to behave as a random variable
is that the sum (\ref{eq:c_n_exact}) consists of many terms with
random phases, and the dynamics of the $c_{n}\left(t\right)$ are
chaotic, since these are generated by the nonlinear Hamitonian, $H_{NLS}$.

A crossover to the regime (\ref{eq:m2-1}) from the regime where a
different power law is found is presented in \cite{Flach2010}.

\subsection{\label{sub:Scaling}Scaling}

The theory presented in the previous subsection assumes that $F_{n}\left(t\right)$
is effectively random inside the relatively flat region of the wavepacket.
Chaotic behavior of the $\psi\left(x\right)$ would generate such
an $F_{n}$. As time evolves, the width of this region increases,
and the density, $\rho$, decreases. The first is expected to enhance
chaos, and the second to suppress it. The calculation of the present
subsection \cite{Pikovsky2011} was performed in order to check which
effect wins. Since one is limited in the length of the system one
can study numerically, the calculation is performed for a system that
is finite but its behavior was found not to change as the size of
the system is increased. Therefore, it can be extrapolated to arbitrarily
large system sizes. In particular, the probability for a system to
be regular as a function of its length, $L$, density, $\rho$ and
the distribution of the random potential, $w$, was studied. A system
is considered regular, if all its Lyapunov exponents vanish, and it
is considered chaotic, if at least one Lyapunov exponent is positive.
In the present subsection a theory for the flat region of Fig. \ref{fig:Shapo}
is presented. For this purpose a dynamics generated by (\ref{eq:NLSE})
for a finite system of size $L$ is examined. The linear system is
invariant under the rescaling,
\begin{equation}
J\to J'=cJ,\qquad w\to w'=cw
\end{equation}
 if time and energy are correspondingly rescaled. In particular, the
localization length is unaffected. For the nonlinear systems also
the rescaling of the norm or the nonlinearity coefficient,
\begin{equation}
\beta\to\beta'=c\beta,
\end{equation}
 is required. In the present work the choice
\begin{equation}
c=\frac{1}{J\left(1+W\right)}
\end{equation}
 was made resulting in,
\begin{equation}
J'=\frac{1}{1+W}\qquad w'=\frac{w}{J\left(1+W\right)}\qquad\beta'=\frac{\beta}{J\left(1+W\right)}.
\end{equation}
 In particular we may choose $J=1$, and $W=w/2$ and study the model
(\ref{eq:NLSE}) with $J$ replaced by $J'=1/\left(1+W\right)$ the
random potential satisfies,
\begin{equation}
-\frac{W}{1+W}\leq\varepsilon_{x}\leq\frac{W}{1+W}
\end{equation}
 and $\beta$ replaced by $\beta'=\beta/\left(1+W\right)$. For convenience
this factor can be absorbed in the definition of the norm of the wavefunction,
\begin{equation}
\mathcal{N}'=\frac{\beta\mathcal{N}}{1+W}.
\end{equation}
 The density is,
\begin{equation}
\rho=\frac{\mathcal{N}'}{L}.
\end{equation}
 It was found that for the fixed $W$ and $\rho$, the probability
to observe regularity decreases with the length $L$. This behavior
can be understood as follows. Suppose we fix $W,\rho$ and consider
a lattice of large length $L$. One divides this lattice into (still
large) subsystems of lengths $L_{0}$. How the probability to observe
regularity on the large lattice $P(\rho,W,L)$ is related to the corresponding
probabilities for smaller lattices $P(\rho,W,L_{0})$? It is reasonable
to assume that to observe regularity in the whole lattice all the
subsystems have to be regular, because any chaotic subsystem will
destroy regularity. This immediately leads the relation
\begin{equation}
P(\rho,W,L)=[P(\rho,W,L_{0})]^{L/L_{0}}\;.\label{eq_scl}
\end{equation}
 Equation (\ref{eq_scl}) implicitly assumes that chaos appears not
due to an interaction between the subsystems, but in each subsystem
(of length $L_{0}$) separately. This appears reasonable if the interaction
between the subsystems is small, i.e. if their lengths are large compared
to the length scale associated with localization in the linear problem:
$L_{0}\gg\xi$ ($\xi$ is the localization length). On these scales
the various subsystems are statistically independent. This is the
content of (\ref{eq_scl}). It motivates the definition of the $L$-
independent quantity,
\begin{equation}
R(\rho,W)=[P(\rho,W,L)]^{1/L}\;.\label{eq_scl2}
\end{equation}
 This scaling relation was verified numerically, starting with a uniform
distribution and using periodic boundary conditions. It was found
that for lattices of sizes $16<L<128$, $R$ is independent of $L$.
Therefore, to calculate $P$ it is sufficient to evaluate,
\begin{equation}
P_{0}\left(\rho,W\right)\equiv P(\rho,W,L_{0}),
\end{equation}
 for a system of some size $L_{0}$. It is assumed that the scaling
found, will hold for systems of arbitrarily large size, which allows
to increase also the localization length. It is convenient to perform
a transformation to a new quantity $Q(\rho,W)$ as $Q=P_{0}/\left(1-P_{0}\right)$,
which yields,
\begin{equation}
P_{0}=\frac{1}{1+Q^{-1}(\rho,W)}.\label{eq:r}
\end{equation}

\begin{figure}[tbh]

\begin{centering}
\includegraphics[width=8cm]{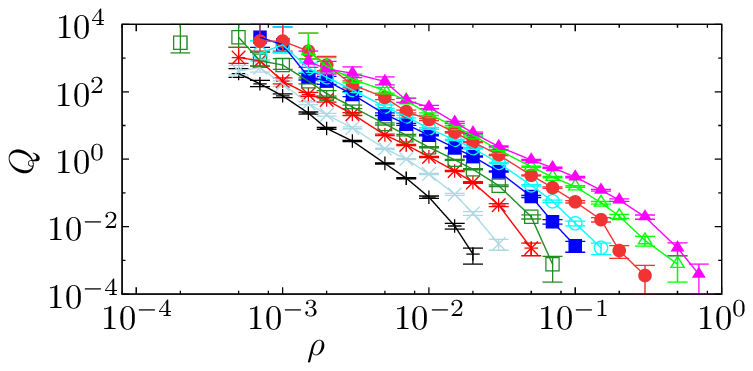}\\
 \includegraphics[width=8cm]{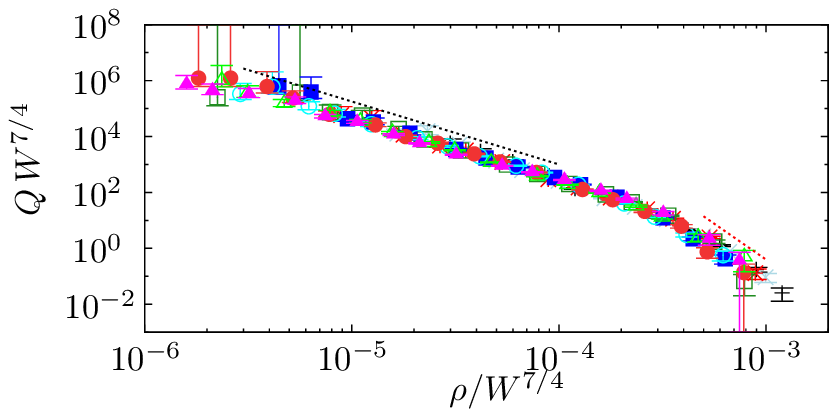}
\par\end{centering}

\caption{\label{fig:scaling}(Color online) The function $Q(\rho,W)$ (top)
and the same function in scaled coordinates (bottom). The black dashed
and red dotted lines, are showing asymptotics for small and large
arguments of $q$, have slopes $\zeta=9/4$ and $\eta=5.2$. (Fig.
4 and 5 of \cite{Pikovsky2011})}
\end{figure}

The limits $\rho\to0$ (linear problem) and $W\to0$ (non-random system)
are singular, therefore it is expected and indeed observed that the
function $Q(\rho,W)$ is not an arbitrary function of $\rho$ and
$W$, but it can be written in a scaling form
\begin{equation}
Q=\frac{1}{W^{\alpha}}q\left(\frac{\rho}{W^{\alpha_{1}}}\right),\label{eq:sc1}
\end{equation}
 where $q(x)$ is a singular function at its limits, $q(x)\sim c_{1}x^{-\zeta}$
for small $x$, while $q(x)\sim c_{2}x^{-\eta}$ for large $x$. The
top of Fig.~\ref{fig:scaling} collapses to one curve as shown in
the bottom of Fig.~\ref{fig:scaling}. This is the numerical justification
for (\ref{eq:sc1}). It also provides the values of the exponents
$\alpha=\alpha_{1}\approx1.75$, $\zeta\approx9/4=2.25$, $\eta\approx5.2$,
$c_{1}\approx2.5\cdot10^{-7}$ and $c_{2}\approx1.8\cdot10^{-18}$.

In particular, for a fixed density, $P_{ch}\sim\rho^{9/4}$, since
the probability of chaos is fixed and non-zero, spreading is expected.
In this aspect model (\ref{eq:NLSE}) differs from the corresponding
$N-$body problem, where for a fixed density and in the thermodynamic
limit, localization was found \cite{Basko2006,Basko2007}. Now let
us assume that we consider the states with the same fixed norm $\mathcal{N}$
on lattices of different length $L$. Then, $\rho=\mathcal{N}/L$,
and one finds
\begin{equation}
P_{ch}\approx\frac{L^{1-\zeta}\mathcal{N}^{\zeta}W^{\alpha(1-\zeta)}}{c_{1}L_{0}}=\frac{L^{-5/4}\mathcal{N}^{9/4}}{c_{1}L_{0}W^{35/16}}.\label{eq_ras1}
\end{equation}
 This quantity, as expected, grows with the norm $\mathcal{N}$ and
decreases with the disorder $W$. We see that because $\zeta>1$,
the probability to observe chaos in large lattices at fixed norm tends
to zero. This result may have implications for the problem of spreading
of an initially localized wave packet in large lattices. In this setup
the norm of the field is conserved, and the effective density decreases
in the course of the spreading. This means that as a function of time
$L$ increases and $P_{ch}$ decreases, therefore, spreading takes
place for less and less realizations of the disorder.

\subsection{\label{sub:Chaotic-spot-model}Study of the relation to phase-space
structures}

It is expected that in the course of spreading, as the nonlinearity
decreases, the trajectories in the $\psi\left(x\right)$, $\psi^{*}\left(x\right)$
phase-space will become more and more regular. In particular, trajectories
that look chaotic on some scale may eventually look regular. For this
purpose the technique of the time-dependent Lyapunov exponents was
introduced \cite{Johansson2010}. It suggests that an initially chaotic
wavepacket may stick to KAM like trajectories resulting in a slow-down
of the spreading. A mechanism for spreading that involved a resonance
of three oscillators resulting in a mechanism for Arnold diffusion
was recently proposed \cite{Basko2011}. It is a spot in phase-space
where the local Lyapunov exponents are positive and there is wandering
in phase-space, presumably, in the regions where tori are destroyed.

A recent qualitative argument for a finite probability for spreading,
combined with numerical results was presented \cite{Ivanchenko2011}.
It is instructive to compare it with rigorous statements on the same
model \cite{Benettin1988}.

\subsection{\label{sub:The-renormalized-perturbation}The renormalized perturbation
theory}

Since there is no spectral theory for nonlinear equations analysis
in the framework of the time dependent perturbation theory was performed
\cite{Fishman2008a,Fishman2009a,Krivolapov2010}. The objective is
to develop a perturbation expansion of the $c_{m}\left(t\right)$
of (\ref{eq:c_n_exact}) in powers of $\beta$ and to calculate them
order by order in $\beta.$ The required expansion is
\begin{equation}
c_{n}\left(t\right)=c_{n}^{\left(0\right)}+\beta c_{n}^{\left(1\right)}+\beta^{2}c_{n}^{\left(2\right)}+\cdots+\beta^{N}c_{n}^{\left(N\right)}+Q_{n},\label{eq:cn_expand}
\end{equation}
 where the expansion is till order $N$ and $Q_{n}$ is the remainder
term (clearly, $c_{n}^{\left(l\right)}$ and $Q_{n}$ are random variables).
The initial condition
\begin{equation}
c_{n}\left(t=0\right)=\delta_{n0},
\end{equation}
 was assumed. The equations for the two leading orders are presented
in what follows. The leading order is
\begin{equation}
c_{n}^{\left(0\right)}=\delta_{n0}.\label{eq:cn0}
\end{equation}
 And the first order is
\begin{equation}
c_{n}^{\left(1\right)}=V_{n}^{000}\left(\frac{1-e^{i\left(E_{n}-E_{0}\right)t}}{E_{n}-E_{0}}\right).\label{eq:1st_order}
\end{equation}
 The divergence of this expansion for any value of $\beta$ may result
from three major problems: the secular terms problem, the entropy
problem (i.e., factorial proliferation of terms), and the small denominators
problem. To eliminate the secular terms the ansatz (\ref{eq:expansion})
is replaced by \cite{Fishman2009a}

\begin{equation}
\psi\left(x,t\right)={\displaystyle \sum\limits _{n}}c_{n}\left(t\right)e^{-iE_{n}^{\prime}t}u_{n}\left(x\right),\label{eq:expansion_prime}
\end{equation}
 where
\begin{equation}
E_{n}^{\prime}\equiv E_{n}^{\left(0\right)}+\beta E_{n}^{\left(1\right)}+\beta^{2}E_{n}^{\left(2\right)}+\cdots\label{eq:En_expand}
\end{equation}
 and $E_{n}^{\left(0\right)}$ are the eigenvalues of $H_{0}$ (clearly,
$E_{n}^{\left(l\right)}$, are random variables), and $E_{n}'$ are
the renormalized energies. The new equation for the $c_{n}$ is given
by

\begin{eqnarray}
i\partial_{t}c_{n} & = & \left(E_{n}^{\left(0\right)}-E_{n}^{\prime}\right)c_{n}\label{eq:diff_eq}\\
 & + & \beta\sum_{m_{1}m_{2}m_{3}}V_{n}^{m_{1}m_{2}m_{3}}c_{m_{1}}^{\ast}c_{m_{2}}c_{m_{3}}e^{i\left(E_{n}^{\prime}+E_{m_{1}}^{\prime}-E_{m_{2}}^{\prime}-E_{m_{3}}^{\prime}\right)t}.\nonumber
\end{eqnarray}
 Inserting expansions (\ref{eq:cn_expand}) and (\ref{eq:En_expand})
into (\ref{eq:diff_eq}) and comparing the powers of $\beta$ \emph{without
expanding the exponent} in $\beta$, produces the following equation
for the $k-th$ order
\begin{eqnarray}
i\partial_{t}c_{n}^{\left(k\right)} & = & -\sum_{s=0}^{k-1}E_{n}^{\left(k-s\right)}c_{n}^{\left(s\right)}+\label{eq:a2}\\
 & + & \sum_{m_{1}m_{2}m_{3}}V_{n}^{m_{1}m_{2}m_{3}}\left[\sum_{r=0}^{k-1}\sum_{s=0}^{k-1-r}\sum_{l=0}^{k-1-r-s}c_{m_{1}}^{\left(r\right)\ast}c_{m_{2}}^{\left(s\right)}c_{m_{3}}^{\left(l\right)}\right]e^{i\left(E_{n}^{\prime}+E_{m_{1}}^{\prime}-E_{m_{2}}^{\prime}-E_{m_{3}}^{\prime}\right)t}.\nonumber
\end{eqnarray}
 The relation to the order by order expansion in powers of $\beta$
was presented in \cite{Krivolapov2010}.

The perturbative approach gives an explicit value of the wavepacket
for long times, with a rigorous control of the error (Fig. \ref{fig:t_star}).
In particular, it rules out spreading for the NLSE with $J=1$, $w=4$,
$\beta=0.1$ for time of order $t=10^{5}$ \cite{Fleishon2011}.

\subsubsection{\label{sec:The-remainder}The remainder of the expansion}

In order to control the solution one has to control, $Q_{n}$, the
remainder of the expansion (\ref{eq:cn_expand}) that can be written
in the form
\begin{equation}
c_{n}\left(t\right)=\tilde{c}_{n}+Q_{n},\label{eq:remainder_def}
\end{equation}
 with
\begin{eqnarray}
\tilde{c}_{n} & = & \sum_{l=0}^{N}\beta^{l}c_{n}^{\left(l\right)}.\label{eq:c_tilde_def}
\end{eqnarray}
 The linear part of the full equation for the remainder is given by
\begin{equation}
i\partial_{t}\tilde{Q}_{n}^{lin}=W_{n}\left(t\right)+\sum_{m}M_{nm}\left(t\right)\tilde{Q}_{m}^{lin}+\sum_{m}\bar{M}_{nm}\left(t\right)\left(\tilde{Q}_{m}^{lin}\right)^{*}.\label{eq:bootstrap_lin}
\end{equation}
 The contribution of the remainder term is
\begin{equation}
\left|Q_{n}\right|\leq\mathrm{const}\cdot e^{6cN^{2}+N\ln\beta+\ln t}e^{-\gamma\left\vert x_{n}\right\vert },\label{eq:remainder_bound}
\end{equation}
 where $\gamma=1/\xi$ is the inverse localization length, $x_{n}$
is the localization center of the $n-$th state. Note that for a given
$t$ and $\beta$ there is an optimum $N$ for which the remainder
is minimal. Additionally, for any fixed time and order $N$, $\lim_{\beta\rightarrow0}\left|Q_{n}\right|/\beta^{N-1}=0$
(see definition at \cite{Fishman2009a}), which shows that the series
is in fact an asymptotic one \cite{Erdelyi1956}. Using a bootstrap
argument, one can show \cite{Fishman2009a} that until some time $t_{*}$
, the dynamics is governed by the linear part and the remainder is
bounded by,
\begin{eqnarray}
\left|Q_{n}\left(t\right)\right| & \leq & A\beta^{N}t\cdot e^{-\gamma\left\vert x_{n}\right\vert },\label{eq:t0_definition}
\end{eqnarray}
 where $\gamma$ is the inverse localization length and $A$ is a
constant. Therefore to estimate the remainder one can integrate (\ref{eq:bootstrap_lin})
at least up to $t_{*}$. It is useful to integrate up to some large
time, $t\ll t_{*},$ and then to extrapolate using the linear bound
(\ref{eq:t0_definition}) up to $t_{*}$. In the next subsection it
will be proposed how to determine $t_{*}$ in practice.

For time shorter than $t_{*}$ there is a front $\bar{x}\left(t\right)\propto\ln t$
such that for $x_{n}>\bar{x}\left(t\right)$ both the remainder, $Q_{n}\left(t\right)$
and $c_{n}\left(t\right)$ are exponentially small.

\subsubsection{\label{sec:Results}Practical estimate of the remainder}

\begin{figure}
\begin{centering}\includegraphics[width=8cm]{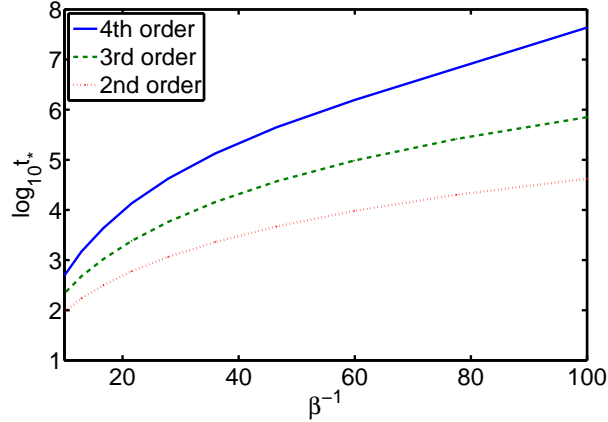}\caption{\label{fig:t_star}$\log_{10}t_{*}$ as a function of $\beta^{-1}$
for different orders. 4th order (solid blue), 3rd order (dashed green)
and 2nd order (dotted red). The parameters are: $w=4$, $J=1$.}
\end{centering}
\end{figure}

In this section it will be demonstrated, how the numerical scheme
for calculations in the framework of the perturbation theory, is implemented
in practice for a specific realization of the random potentials $\left\{ \varepsilon_{x}\right\} $.
It was found that there are some modes that result in the largest
contributions to the remainder. These modes are located near the initial
mode and therefore can be easily identified and subtracted. For numerical
evaluations one defines a norm
\begin{equation}
\left\Vert Q\right\Vert _{2}\equiv\left(\sum_{m}\left|Q_{m}\right|^{2}\right)^{1/2},
\end{equation}
 the definition of $\left\Vert Q_{lin}\right\Vert _{2}$ is similar.
It was found that for $\left\Vert Q_{lin}\right\Vert _{2}\leq0.1$,
$\left\Vert Q_{lin}\right\Vert _{2}$ is close to $\left\Vert Q\right\Vert _{2}$
, which leads to the definition of $t_{*}$,
\begin{equation}
\left\Vert Q_{lin}\left(t_{*}\right)\right\Vert _{2}=0.1.\label{eq:t_star_def}
\end{equation}
 For small nonlinearity strength, $\beta$, $t_{*}$ is very large
and therefore the integration of (\ref{eq:bootstrap_lin}) to $t_{*}$
is very time consuming. Equation (\ref{eq:t_star_def}) can be used
to \emph{extrapolate linearly} from the time interval where (\ref{eq:bootstrap_lin})
is solved to $t_{*}$. Practically, one can find $t_{*}$ from (\ref{eq:t_star_def})
by extrapolation. In Fig. \ref{fig:t_star} a plot of $\log_{10}t_{*}$
as a function of $\beta^{-1}$ for different orders is presented.

A systematic improvement with the order of the perturbation theory
is found. Note, that even with moderate nonlinearity strengths, namely,
$\beta<0.08$, one can achieve a good approximation of the solution
up to very large times. In this way the perturbation theory combined
with the solution of the linear equation (\ref{eq:bootstrap_lin})
and the criterion (\ref{eq:t_star_def}) may be used to obtain the
solution of the original equation (\ref{eq:NLSE}) up to $t<t_{*}$.
For small $\beta$ the time $t_{*}$ is very long, as is clear from
Fig. \ref{fig:t_star}. Removing the dominant modes, which were mentioned
in the beginning of this subsection, allows to obtain reliable results
for times larger by more than one order of magnitude from the times
presented at Fig. \ref{fig:t_star} (see, Fig. 10 in \cite{Krivolapov2010}).
When one considers the smallness of $\beta$ one should consider actually
the smallness of $\beta e^{2}$ due to the exponential proliferation
of the number of terms (see Eq. (4.6) of \cite{Fishman2009a}).

\section{\label{sec:Rigorous-results}Rigorous results}

The rigorous analysis of the NLSE equation with a random potential
turned out to be very difficult as well. The results so far are very
limited in scope, yet of sufficient interest to point out the nature
of the problem at hand. Most notably we have the following conclusive
results.

\subsection{\label{sub:Many-body-localization}Many body localization}

Consider the system of $N$ interacting particles via finite range
interactions, on a finite $d-$ dimensional lattice, $\mathbb{Z}^{d}$,
with the Hamiltonian
\begin{equation}
H^{\left(N\right)}\left(\omega\right)=\sum_{j=1}^{N}\left[-\Delta_{j}+\lambda V\left(x_{j};\omega\right)\right]+\sum_{i<j}V_{ij}\left(x_{i}-x_{j}\right)
\end{equation}
 acting on the Hilbert space $\mathcal{H}^{\left(N\right)}=\ell^{2}\left(\mathbb{Z}^{d}\right)^{N}.$
It is assumed, for simplicity, that the $V_{ij}$ are all uniformly
bounded by a constant, $0<\alpha<\infty$, and the random potential
$V\left(x_{j};\omega\right)$ is given in terms of a collection of
i.i.d random variables, $\left\{ V\left(x,\omega\right)\right\} _{x\in\mathbb{Z}^{d}}$,
with bounded probability distribution and finite moment generating
function
\begin{equation}
\mathbb{E}\left(\exp\left(tV\left(0\right)\right)\right)<\infty\qquad\forall t.
\end{equation}
 Under the above assumptions on the potentials, it was proved that
for a large disorder, the spectrum of $H^{\left(N\right)}$is a dense
point spectrum with probability 1, and the eigenfunctions are exponentially
localized, after an appropriate distance function is defined on the
lattice, between clusters of particles \cite{Aizenman2009,Chulaevsky2009}.
This important result shows that the inter-particle interactions between,
say, electrons in a solid, cannot destroy Anderson localization, at
least under some favorable situations. However, this result requires
the size of the disorder to be dependent on $N$, the number of particles!
Therefore, it cannot be applied to systems with non-zero density of
particles, as systems studied in \cite{Basko2006,Basko2007}.

\subsection{\label{sub:Quasi-periodic-perturbations}Quasi-periodic perturbations}

We now turn to the fully nonlinear problem. The first important result
in this case is \cite{Frohlich1986}, where the time-independent problem
was explored. The nonlinear term of the NLSE is considered as a small
perturbation of the linear dynamical system corresponding to the linear
Anderson problem on the lattice. One is then led to consider the KAM
theory in infinite dimensional phase--space, as a way to construct
periodic and quasi-periodic solutions (in time) for such models. This
has been shown to apply to models with good Diophantine properties
(linear combination with integer coefficients, bounded away from zero)
of the eigenvalues of the linear part. The possibility that the solutions
are localized in space and quasi-periodic in time, notably leads to
the question of whether linearizing the NLSE around such a solution,
will result in a linear system with localized states only. The idea
behind the use of quasi-periodic in time models is that it comes from
a formal iterative scheme for solving the NLSE with localized solutions.
If a localized solution of the equation is assumed, it can be expanded
in terms of the normal modes (eigenfunctions) of the linear problem,
so that
\begin{equation}
\psi_{N}\left(x\right)=\sum_{j=1}^{b\left(N\right)}c_{j}u_{j}\left(x\right)e^{-iE_{j}t},
\end{equation}
 where $N$ is the iteration number, and $b\left(N\right)$ is the
number of modes generated by $N$ iterations. In this approximation
the nonlinearity is given by,
\begin{equation}
\lambda\left|\psi_{N}\right|^{2}=\lambda\left|\sum_{j=1}^{b\left(N\right)}c_{j}u_{j}\left(x\right)e^{-iE_{j}t}\right|^{2}.
\end{equation}
 Therefore, the approximate dynamics is governed, to this order by,
\begin{equation}
i\partial_{t}\psi=\left(-\Delta+V\right)\psi+\lambda\left|\psi_{N}\right|^{2}\psi.
\end{equation}
 A uniform (in $N)$ proof of complete dynamical localization, for
the linear equation presented above, would imply localization for
the NLSE problem. However, in this iteration scheme $b\left(N\right)$
grows very fast in $N$, and furthermore, the frequencies in $\lambda\left|\psi_{N}\right|^{2}$
may be of an arbitrarily small value, which is hard to control \cite{Bourgain2004}.
This is just another manifestation of the small denominator problem
one encounters in other perturbative methods.

This approach was followed in a series of papers, beginning with \cite{Soffer2003},
where the Anderson model on a lattice is perturbed by an exponentially
localized time-periodic potential. It was shown that the corresponding
Floquet operator has purely dense point spectrum, with exponentially
localized eigenfunctions. This implies, by general spectral theory
that the original equation has dynamical localization, namely, the
initial solution does not spread. This was further generalized to
the much harder case of exponentially localized, quasi-periodic in
time potential perturbations of the Anderson model on the lattice
with a similar result - complete localization for small potential
perturbations \cite{Bourgain2004}. The key estimates can be formulated
as localization for the Floquet operator of the type (for two non-commensurate
frequencies)
\begin{equation}
K\equiv-i\omega_{1}\frac{\partial}{\partial\theta_{1}}-i\omega_{2}\frac{\partial}{\partial\theta_{2}}+\epsilon\Delta+V+W_{1}\left(j\right)\cos2\pi\theta_{1}+W_{2}\left(j\right)\cos2\pi\theta_{2},
\end{equation}
 acting on the Hilbert space $H_{K}\equiv\ell^{2}\left(\mathbb{Z}^{d}\right)\otimes L^{2}\left(\mathrm{T}^{2}\right)$
(here, $\mathrm{T}^{2}$ stands for a two-dimensional torus). Then
localization was proved with large probability, and for a corresponding
set of Diophantine conditions needed in the quasi-periodic case \cite{Soffer2003,Bourgain2004}.
In particular, it was shown that with large probability, $K$ has
a dense point spectrum, with exponentially decaying eigenfunctions.
However, the estimates deteriorate as the quasi-periodic frequencies
approach zero, which makes them hard to use for the solution of the
full NLSE problem.

Further results on models approximating in various ways the NLSE problem,
include \cite{Bourgain2008}, where it is shown that quasi-periodic
solutions in time exist, and that they bifurcate from the corresponding
quasi-periodic solutions of the linear equation. Another result is
that if the nonlinearity coupling constant vanishes at space infinity
(at some polynomial rate), then the NLSE with Anderson potential on
the lattice has only localized solutions for large disorder \cite{Bourgain2007}.

\subsection{NLSE with a random potential}

Turning our attention to the full NLSE with a random potential problem,
very little is known. The most important result, which indicates localization
for large times with large disorder and small nonlinearity is in the
work of \cite{Wang2008}. The main result can be formulated as follows.
The small parameter in the problem is $\epsilon=J+\beta$, and $w=1$.
Therefore, small $\epsilon$ implies strong disorder $\left(J\ll1\right)$
and weak nonlinearity $\left(\beta\ll1\right)$. For an initial data
in $\ell^{2},$ localized at the origin, in the sense that,
\begin{equation}
\sum_{\left|j\right|>R}\left|\psi\left(j,t=0\right)\right|^{2}\leq\delta\ll1,
\end{equation}
 it was proven\cite{Wang2008}, that for all $A$ and $\delta>0$
there exists a constant $C\left(A\right)>0,$ $\epsilon\left(A\right)>0$
and $K\left(A\right)>A^{2}$ such that for all $t\leq\left(\delta/C\right)\epsilon^{-A}$
,
\begin{equation}
\sum_{\left|j\right|>R+K}\left|\psi\left(j,t\right)\right|^{2}\leq2\delta,
\end{equation}
 with probability
\begin{equation}
1-\exp\left(-\frac{R}{K}e^{-2K_{\epsilon}^{1/CA}}\right),
\end{equation}
 for all $0<\epsilon<\epsilon\left(A\right)$. In this context, it
should be noted that the methods used to prove such results, follow
the infinite dimensional generalization of dynamical systems theory.
It includes a construction of normal form transformations to approximately
{}``diagonalize'' the system. Such methods were used already in
\cite{Benettin1988} to prove a similar result for the nonlinearly
coupled random masses. This result implies that the $\ell^{2}-$norm
grows at most logarithmically in time \cite{Wang2008}. However, it
is hard to compare these estimates to other results, since the constants
are not controlled in this analysis. Note, that the corresponding
results (\ref{eq:remainder_bound}) and (\ref{eq:t0_definition})
of the perturbation theory are for arbitrary strength of disorder,
and improve when it increases.

Another class of results, which give explicit estimates on the solution
of NLSE with a random potential was developed in \cite{Fishman2008a,Fishman2009a,Krivolapov2010}.
One can construct renormalized time-dependent perturbation theory
for the solutions, starting from the eigenfunctions of the linear
problem as a basis. It is then shown, that after eliminating the secular
terms, we get an explicit expansion, which is computable. Each term
depends explicitly on the potential of the linear problem and the
eigenvalues. Therefore, by using the known and some new results on
the \emph{linear }spectral problem it is possible to estimate the
various terms in the expansion on average. In particular, in \cite{Rivkind2011}
it is shown that for the linear problem in one dimension, if the potentials
are uniformly bounded by some finite constant, then for \emph{all}
potentials the minimal distance between the eigenvalues is nonzero,
bounded below by $e^{-cN}$, where $N$ is the size of the lattice.
This is then used to control rigorously the first order term in the
above expansion (\ref{eq:cn_expand}),(\ref{eq:1st_order}), and to
conclude that to this order the exponential localization persists
in the nonlinear case. It should be noted that the lower bound on
the distance between the eigenvalues, though exponentially small,
is sufficient to the purpose of controlling the various terms in the
renormalized expansion. It also allows to obtain bounds without the
usual Diophantine conditions. Moreover, it implies a bound on the
expectation of the derivative of the eigenfunctions with respect to
the potential,
\begin{equation}
\sum_{i=1}^{N}\mathbb{E}\left|\frac{\partial\psi_{i}\left(x\right)}{\partial\varepsilon_{j}}\right|\leq CN.
\end{equation}
 However, to control the higher order terms in the expansion, a control
of linear combinations of more than two eigenvalues is needed. In
the work of \cite{Aizenman2008} there are Diophantine estimates on
the eigenvalues of the linear problem that can similarly control many
other terms in the expansion of \cite{Fishman2008a,Fishman2009a,Krivolapov2010}.
However, this is not sufficient for complete control in the probabilistic
sense of the linear combinations of the eigenvalues of the linear
problem, to bound all the terms in the expansion.

\section{\label{sec:Numerics}Numerical results}

The main problem one encounters with the numerical calculation for
chaotic systems is the exponential sensitivity to numerical errors.
For this reason one cannot perform the calculation for substantially
long time scale with a control on the errors of a specific solution.
The validity of the results is traditionally tested by changing the
size of the steps verifying that the results are not affected. The
correct result is assumed to be the one that is found in the limit
of a vanishing step size. The main problem with such assumptions is
that the limit of zero time step may be singular. Moreover, there
is no theoretical understanding that the numerical solutions are some
types of sampling of the phase-space as is the situation for chaotic
systems. Typically, one finds that after some time, spreading starts,
as shown in Fig. \ref{fig1}. Eventually, it turns into a sub-diffusion
and the second moment grows according to (\ref{eq:m2-1}) with the
exponent $\nicefrac{1}{3}$ \cite{Flach2009,Flach2009a,Skokos2009,Laptyeva2010,Bodyfelt2011}.
The wavepacket is typically of the shape presented in Fig. \ref{fig:Shapo}.
A theory similar to the one presented in Subsection \ref{sub:Effective-noise-theories}
but with the nonlinear term (\ref{eq:Gen_Nonlin_term}) was developed
and was found to agree with the numerical results \cite{Pikovsky2008,Skokos2009,Flach2009,Mulansky2009,Skokos2010}.
The longest time for which numerical calculations were performed is
$t=10^{8}$ (in units where $J=1$ in (\ref{eq:NLSE})).

In order to follow the dynamics of a wavepacket, typically the SABA
algorithm is used. This algorithm belongs to the family of split step
algorithms and evaluates the wave packet in small steps, changing
from coordinate space to momentum space. The disorder and nonlinear
interaction are applied in the coordinate space, the wave is then
transformed to the momentum space, where the kinetic energy term is
applied. The solution in real space is recovered by transforming it
back to the coordinate space, and the procedure is repeated. Nearly
all numerical calculations for this problem use such methods. Additional
details on the SABA algorithm, can be found in \cite{Skokos2009}.
Like any numerical algorithm, the SABA algorithm accumulates errors
during the calculation which grow with the time of the integration.
There is no reliable estimate of these errors.

Double humped states were studied numerically in the presence of nonlinearity
that is not too strong. It was found that the spreading of a wave
packet prepared initially near some site $O$ is substantially stronger
if there is a double humped state with one of its humps near $O$,
than if the states peaked near $O$ are single humped. It was found
\cite{Veksler2010} that there is a regime of parameters where $\beta$
is sufficiently small so that the double humped structure is preserved
but the packet is not only oscillating between the humps but also
leaks to other states, leading to spreading. Additionally, if $\beta$
is small enough so that the oscillations between the two states are
not suppressed in the double-well model, then the double humped states
will contribute to the spreading for the NLSE. But since double humped
states are suppressed and do not contribute to the spreading for high
nonlinearities, it cannot be claimed that they dominate the spreading
for the NLSE.

\section{Discussion and open problems}

All numerical calculations exhibit spreading of an initially localized
wavepacket (although some part of it may not spread). The spreading
results in sub-diffusion with the exponent $\nicefrac{1}{3}$. All
rigorous and analytical theories predict that asymptotically the spreading
cannot be faster than logarithmic in time. The main difficulty is,
that there is no regime of parameters, where analytical and numerical
results agree for a long time. For short times $\left(t\leq10^{4}\right)$,
perturbation theory was found to agree with the numerical results
(and for $t\leq10^{5}$ in \cite{Fleishon2011}).

\subsection{List of open problems}

We list the questions that may be explored by the various communities:
\begin{enumerate}
\item What is the asymptotic time scale? Namely, what is the time scale
where the theories predicting suppression of sub-diffusion become
effective.
\item When does spreading start and how this time depends on parameters?
\item Is it possible to analytically derive the scaling theory presented
in Subsection \ref{sub:Scaling}?
\item Can one prove the bound of the terms of the perturbation theory presented
in Subsection \ref{sub:The-renormalized-perturbation}? In this theory
it is difficult to control denominators of the form $f_{l}=\sum_{i=1}^{l}c_{i}E_{i}$
where $E_{i}$ are the eigenvalues of the Anderson Hamiltonian, $H_{0}$,
and $c_{i}$ are integer constants. Numerical calculations show that
these satisfy a central limit theorem. In particular, one would like
to prove that,
\begin{equation}
\left\langle \frac{1}{\left|f_{l}\right|^{s}}\right\rangle <\infty\qquad\text{for }0<s<1.
\end{equation}

\item Is the system chaotic? Namely, is there an exponential instability
of the motion in the $\psi\left(x\right)$, $\psi^{*}\left(x\right)$
phase-space? This is a fundamental question since this phase-space
is of infinite dimension.
\item Are there KAM tori? Is there sticking to these tori and is it stable
to the numerical errors?
\item Can one design experiments that can be extended to the asymptotic
long time regime?
\item May the control of the numerical scheme be improved?
\item Another possible mechanism for spreading is tunneling to exponentially
large distances with exponentially small probability in space and
time. This cannot be decided using numerical calculations, since only
relatively small systems are accessible. Another method should be
developed to resolve this issue.
\end{enumerate}

\ack{}{}

We would like to thank S. Flach, D. Krimer and A. Pikovsky for providing
detailed information and useful comments required to the review. This
work was partly supported by the Israel Science Foundation (ISF),
by the US-Israel Binational Science Foundation (BSF), by the USA National
Science Foundation (NSF DMS-0903651), by the Minerva Center of Nonlinear
Physics of Complex Systems, and by the Shlomo Kaplansky academic chair.
AS thanks M. Segev and SF for the hospitality at the Technion, where
this review was prepared.

\section*{References}

 \bibliographystyle{unsrt}
\bibliography{NLSE}

\end{document}